\documentclass{article}
\usepackage{epsfig}
\usepackage{rotating}
\usepackage{exscale}
\usepackage{amsmath}
\usepackage{cite}
\usepackage{latexsym}
\usepackage{graphics,color}
\usepackage[english]{babel}
\newcommand{\bm}{\begin{minipage}}
\renewcommand{\em}{\end{minipage}}
\newcommand{\bfi}{\begin{figure}}
\newcommand{\efi}{\end{figure}}
\newcommand{\be}{\begin{eqnarray}}
\newcommand{\en}{\end{eqnarray}}
\newcommand{\bc}{\begin{center}}
\newcommand{\ec}{\end{center}}
\newcommand{\im}{\mbox{Im}}
\newcommand{\re}{\mbox{Re}}
\begin{document}
\title{A description of the ratio between electric and 
magnetic proton form factors
by using space-like, time-like data and dispersion relations
}
\date{}
\author{R. Baldini$^{\rm a,b}$,
        C. Bini$^{\rm d}$,
        P. Gauzzi$^{\rm d}$, M. Mirazita$^{\rm b}$,
        M. Negrini$^{\rm c}$, S. Pacetti$^{\rm b}$
\vspace{10mm}\\
$^{\rm a}$ Centro Studi e Ricerche Enrico Fermi, Roma, Italy\\
$^{\rm b}$ Laboratori Nazionali di Frascati dell'INFN, Frascati, Italy\\
$^{\rm c}$ Universit\`a di Ferrara e Sezione INFN, Ferrara, Italy\\
$^{\rm d}$ Universit\`a ``La Sapienza'' e Sezione INFN, Roma, Italy
}
\maketitle
\baselineskip=11.6pt
\begin{abstract}
We use the available information on the ratio between
the electric and magnetic proton form factors coming from recently
published space-like data and from the few available time-like data.
We apply a dispersive procedure on these data to evaluate the behaviour 
of this ratio, as a complex function, for all values of $q^2$.
\end{abstract}
%
%\baselineskip=14pt
%

%%%%%%%%%%%%%%%%%%%%%%%%%%%%%%%%%%%%%%%%%%%%%%%%%%%%%%%%%%%%%%%%%%%%%%%%%

\section{Introduction}
\label{sec:introduction}

The electromagnetic form factors (ff) are essential pieces of our 
knowledge of the internal structure of the nucleon, and this 
fact justifies the efforts devoted to their determination.
They are complex functions of the squared momentum transfer 
in the photon-nucleon vertex, defined both for space-like 
($q^2 < 0$) and time-like ($q^2 > 0$) momenta.
The interest in the study of the nucleon ff's has been 
recently renewed \cite{brodsky,arr_int} by the Jlab unexpected 
results on the ratio
\be
R(q^2)=\mu_p\frac{G_E^p(q^2)}{G_M^p(q^2)}, 
\label{ratio}
\en
where $\mu_p$ is the proton magnetic moment and $G_E^p(q^2)$ and
$G_M^p(q^2)$ are the electric and magnetic proton ff's.
A decrease of $R(q^2)$ as space-like $|q^2|$ 
increases \cite{jlab} has been found, in contrast with the flat behaviour 
$R(q^2) \approx 1$ \cite{sl_data} obtained up to 
$q^2 \approx -7\;GeV^2$ by using the traditional Rosembluth 
method~\cite{rosemb}.
It has to be mentioned that
the IJL nucleon ff's model \cite{iachello}, somehow 
related to soliton models of the nucleon \cite{soliton}, predicted the 
decrease of $R(q^2)$ more than thirty years ago.
Recently this model \cite{IJL2} and other QCD based models 
\cite{brodsky,1/q,logq2} have been extrapolated to time-like positive $q^2$. 
While they agree within the experimental errors in the
space-like region, they disagree in some respect 
for time-like $q^2$.
In all these models the space-like $q^2$ variation is 
related to a cancellation between the Dirac, $F_1^p(q^2)$, and 
the Pauli, $F_2^p(q^2)$, ff's. Then, since we have:
\be
\begin{array}{l}
G_E^p(q^2)=F_1^p(q)+\tau F_2^p(q^2)\\
\\
G_M^p(q^2)=F_1^p(q)+ F_2^p(q^2)\\
\end{array}\hspace{10mm} \tau=\frac{q^2}{4M_p^2},
\label{dirac-pauli}
\en
such a cancellation should become an enhancement once $q^2$ 
(i.e. $\tau$) changes sign from negative to positive, time-like values.
As a consequence the angular distribution of the outgoing nucleon
in $e^+e^-\rightarrow N\overline{N}$ processes should have a large $\sin^2(\theta)$
term proportional to $|R(q^2)|^2$.
At the same time, a large transverse polarization of the outgoing 
nucleon \cite{dubn} is predicted \cite{brodsky}, but the sign and 
the $q^2$-dependence strongly depends on the models.
\par\noindent

In principle, time-like ff's could be evaluated from the
space-like ones by means of dispersion relations (DR), if they are smooth
enough and if space-like data were known with very high accuracy or 
if in the time-like region there are some data or suitable 
constraints \cite{gourdin,baldini,attempts}.  
%
%\textcolor{red}{
%Hence, for over 40 years, many attempts have been made 
%to unravel the nucleon ff's in the space-like and time-like regions,
%by applying the DR's to the electric or magnetic ff\cite{baldini,attempts}, 
%but never directly to their ratio.}
%
In fact, analytic functions are supposed to describe at the same time
space-like and time-like electric and magnetic ff's.
These functions are defined in the whole $q^2$ complex plane;
in the physical sheet they do not have isolated singularities, such as poles,
but a cut on the real axis starting at $s_0= (2 m_{\pi})^2$.
Cauchy theorem allows to relate the space-like real values of a 
ff to an integral, over the time-like cut, of its 
imaginary part, providing a DR among them. 
In the unphysical region, $0 < q^2 < 4 M^2_{p}$, not directly
accessible experimentally, each ff should have large bumps
corresponding to the unphysical excitations of $\rho$, $\omega$, $\rho'$,
$\omega'$ and all vector mesons whose mass is lower than $2 M_{p}$ \cite{vmd}.
In terms of ff's, these resonances are described by poles 
lying in the unphysical sheet of the $q^2$ complex plane. 
In the ratio $R(q^2)$ the effects of these poles should be somewhat 
smoothed out, being the same in both, $G_{E}^p$ and $G_{M}^p$. 
Hence $R(q^2)$ smoothness should be a plausible ansatz.\\
There are constraints to $R(q^2)$ for time-like $q^2$, namely:
\begin{itemize}
\item  a continuous transition is expected from space-like to time-like
$q^2$; 
\item according to the Phragm\`en-Lindel\"off theorem the space-like 
and time-like values of a ff must be asymptotically equal in 
modulus\cite{math,asymp1,asymp2};
\item at $q^2=s_1\equiv 4M_p^2$, the physical threshold of $e^{+}e^{-} \rightarrow p \bar{p}$,
assuming $F_1^p(q^2)$ and $F_2^p(q^2)$ analytic functions of $q^2$, it is
$R(s_1) =\mu_p$. Since in this $q^2$ region $R(q^2)$ is
a complex function, this is a constraint on both, real and imaginary
part.
\end{itemize}

All the published data in the time-like region assumed $|R(q^2)|=\mu_p$ 
not only at threshold \cite{adone,dm1,dm2_1,dm2_2,apple,fenice_1,fenice_2},
 but in the whole explored interval \cite{e760,e835_1,e835_2}, essentially 
for lack of accurate data concerning the angular distributions and 
of any measurement of the outgoing nucleon polarization or any use 
of polarized beams. 

In this paper, the ratio $R(q^2)$ has been obtained in a model independent 
way, assuming it has a smooth behaviour, taking into account the theoretical 
constraints and solving the DR by a minimization algorithm.
The input of this algorithm is given by the JLab and MIT-Bates 
polarization data\cite{mit,jlab} in the space-like region 
and by the result of a reanalysis of FENICE\cite{fenice_1,fenice_2}, 
DM2\cite{dm2_1,dm2_2} and E835\cite{andreotti}
data in the time-like region. A preliminary version\cite{b2004} of this 
paper has been previously presented. 

A very similar approach, even if with a different dispersive integral,
has already been successfully tested on the pion ff
\cite{baldini}. In short, it has been assumed the pion ff to be
known in the space-like region and in the time-like region, above
$s_1$ only. Hence the DR has been reversed evaluating the time-like pion ff 
below $s_1$
by means of a standard regularization procedure used to solve first kind 
integral equations, that requires one free parameter only \cite{unfolding}. 
A very good agreement has been obtained in this way, varying this 
parameter within one order of magnitude. This procedure has also been 
applied quite successfully to get the magnetic nucleon ff's in the 
unphysical region \cite{baldini}.

%%%%%%%%%%%%%%%%%%%%%%%%%%%%%%%%%%%%%%%%%%%%%%%%%%%%%%%%%%%%%%%%%%%%%%%%%%%%%%%%%%%%%%%%%%%%%%%%%%%%%%

\section{Space-like and time-like experimental data}
\label{sec:sl_data}

The high intensity-high polarization electron beams available at JLab 
and MIT-Bates\cite{mit,jlab} allowed the extraction of the ratio 
between electric and magnetic proton ff's measuring the 
electron-to-proton polarization transfer.
These measurements showed an almost linear decrease of $R(Q^2)$ from unity
at low $Q^2\equiv-q^2$ up to $\approx 0.3$ at the highest $Q^2$, as shown in 
Fig.\ref{data-fig}, in strong disagreement with previous Rosembluth 
measurements (see ref.\cite{sl_data} and ref.\cite{arr_ana} for a
compilation of all the space-like data), that indicated approximate 
ff scaling, i.e. $R(Q^2) \approx 1$, 
though with large uncertainties in $G^p_{E}$ at the highest $Q^2$ values.
In the Rosembluth measurements, the ff's are basically extracted
fitting the linear $\epsilon$-dependence of the cross section at fixed
$Q^2$ (where $\epsilon$ is the virtual photon polarization), but the
presence of the factor $1/\tau$ [see eq.(\ref{dirac-pauli})] in front of 
$G_E^p(Q^2)$ makes difficult its extraction when $Q^2$ becomes large.
Then, possible experimental inconsistency could affect the Rosembluth
result, however a recent reanalysis \cite{arr_ana} showed that the
individual Rosembluth measurements are consistent to each other within
small normalization uncertainty between the different experiments.
In addition, the new Rosembluth measurements performed at JLab
\cite{super-ros} confirmed the scaling behaviour of the old data for $Q^2$
from 2.6 to 4.1 $GeV^2$, making it clear that the source of the
discrepancy is not simply experimental.
\bfi[h!]\vspace{-5mm}
\bc
\epsfig{file=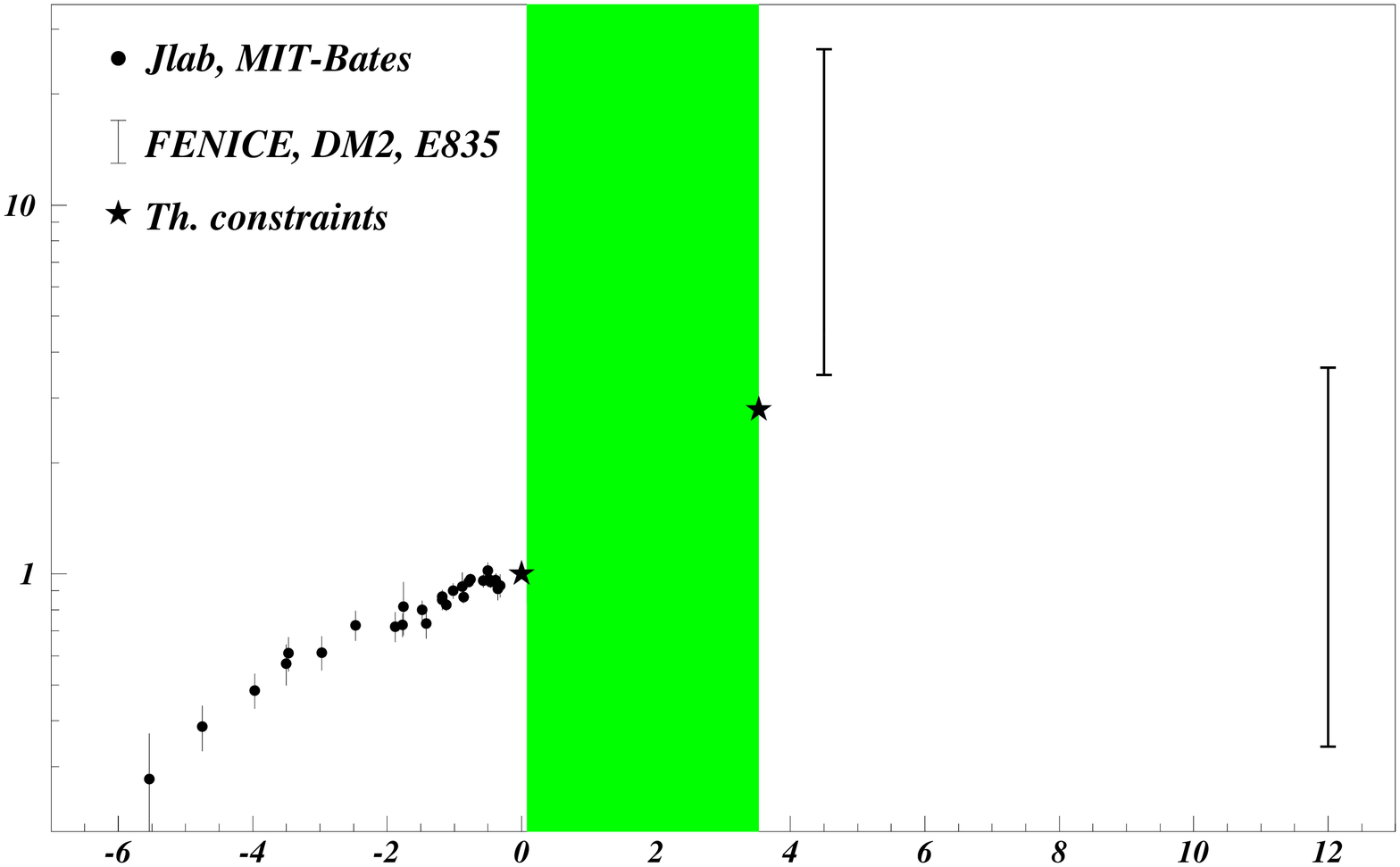,width=120mm}
\put(-65,5){$q^2(GeV^2)$}
\put(-335,181){\rotatebox{90}{$\tilde{R}(q^2)$}}
\caption{The full circles represent the data from Jlab and 
MIT-Bates\cite{mit,jlab}, while the two
time-like intervals are from FENICE\cite{fenice_1,fenice_2}, 
DM2\cite{dm2_1,dm2_2} and E835\cite{andreotti}. The stars are the
theoretical constraints and the shadowed area is the unphysical
region. The label $\tilde{R}(q^2)$ of the ordinate axis is defined as:
$\tilde{R}(q^2)=R(q^2)$ for $q^2\le s_0$ and $\tilde{R}(q^2)=|R(q^2)|$ for 
$q^2> s_0$.}
\label{data-fig}
\ec
\efi\vspace{-0mm}\\
Recent theoretical works \cite{2pho} have suggested that terms,
related to two-photon exchange corrections to the lowest order QED
diagram, may result in incorrect determination of the ff's from
the measured cross section, while polarization measurements should be less
sensitive to such corrections.
As shown in ref.\cite{afan}, the Rosembluth data with the inclusion of 
two-photon exchange contribution agree well with polarization data for $Q^2$ 
between 2 and 3 $GeV^2$, while there is at least partial reconciliation 
for higher $Q^2$. Once reliable calculations of these corrections will be 
available, then polarization and cross section data can be consistently 
combined to extract the ff's without ambiguity.

In our following analysis, we will make use of the ratio $R(Q^2)$ 
as obtained from polarization measurements, since, as discussed,
these data are less sensitive to higher order corrections and to
systematic uncertainties.

%%%%%%%%%%%%%%%%%%%%%%%%%%%%%%%%%%%%%%%%%%%%%%%%%%%%%%%%%%%%%%%%%%%%%%%%%%%%%%%%%%%%%%%%%%%%%%%%%%%%%%

\begin{figure}[h!]\vspace{-5mm}
\begin{center}
\epsfig{file=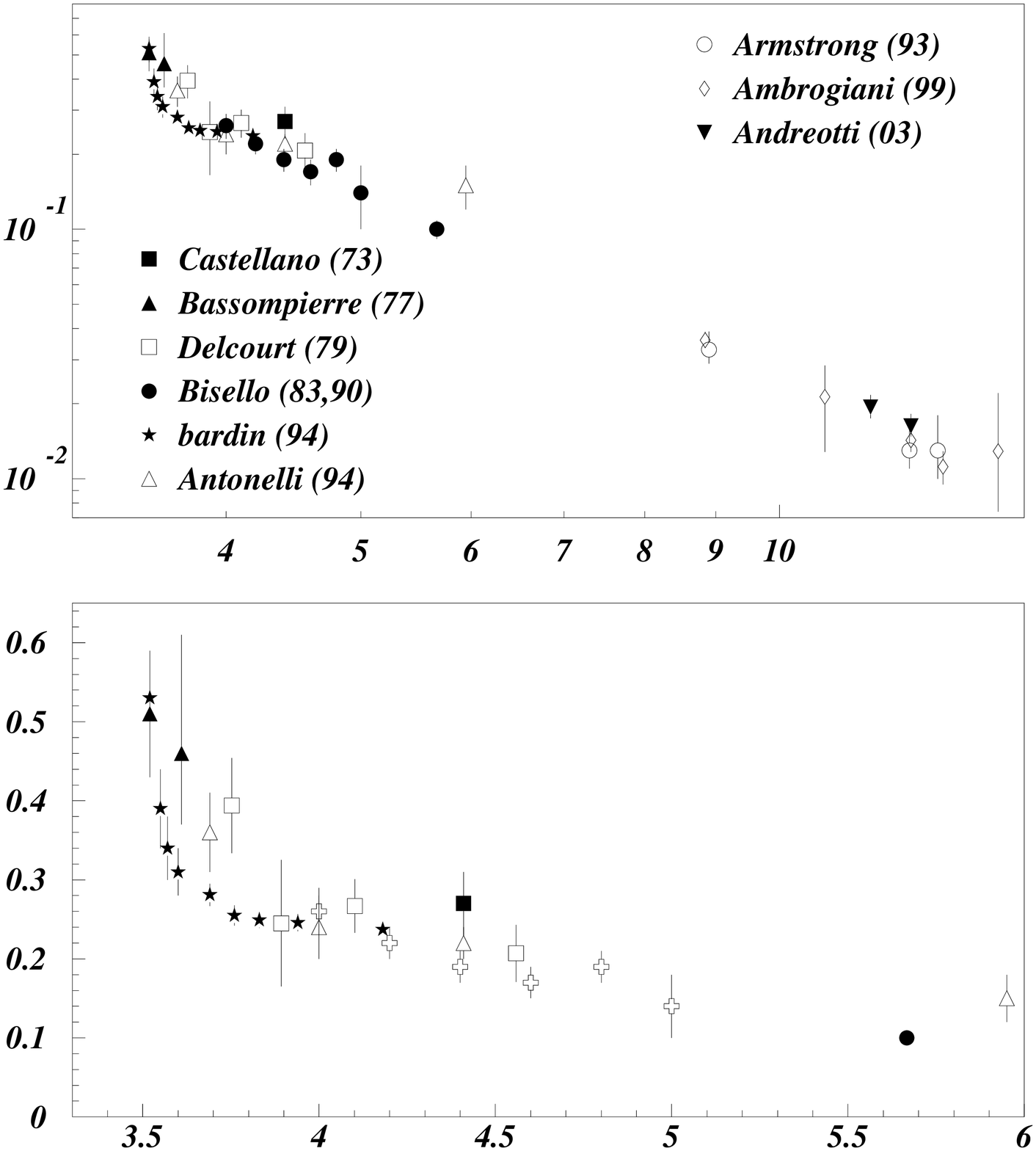,width=100mm}
\put(-60,160){$s(GeV^2)$}
\put(-60,5){$s(GeV^2)$}
\put(-280,269){\rotatebox{90}{$|G_M^p(s)|$}}
\put(-295,118){\rotatebox{90}{$|G_M^p(s)|$}}
\put(-185,235){\scalebox{.8}{\cite{adone}}}
\put(-174,224){\scalebox{.8}{\cite{dm1}}}
\put(-192,213){\scalebox{.8}{\cite{dm2_1}}}
\put(-188,202){\scalebox{.8}{\cite{dm2_2}}}
\put(-199,191){\scalebox{.8}{\cite{apple}}}
\put(-190,180){\scalebox{.8}{\cite{fenice_1,fenice_2}}}
\put(-44,289){\scalebox{.8}{\cite{e760}}}
\put(-40,278){\scalebox{.8}{\cite{e835_1}}}
\put(-48,267){\scalebox{.8}{\cite{e835_2}}}
\put(-256,294){a}
\put(-256,140){b}
\vspace{-0mm}
\caption{Proton magnetic ff in the time-like region 
extracted from the $e^+e^-\rightarrow p\overline{p}$ and
$p\overline{p}\rightarrow e^+e^-$ cross-sections assuming 
$|G^p_E(s)|=|G^p_M(s)|$. a) All data including the largest $s$
ones in logarithmic scale. b) Data close to $s_1=4M_p^2$ in
linear scale.}
\label{collection}
\end{center}
\end{figure}

\begin{figure}[h!]\vspace{-5mm}
\begin{center}
\epsfig{file=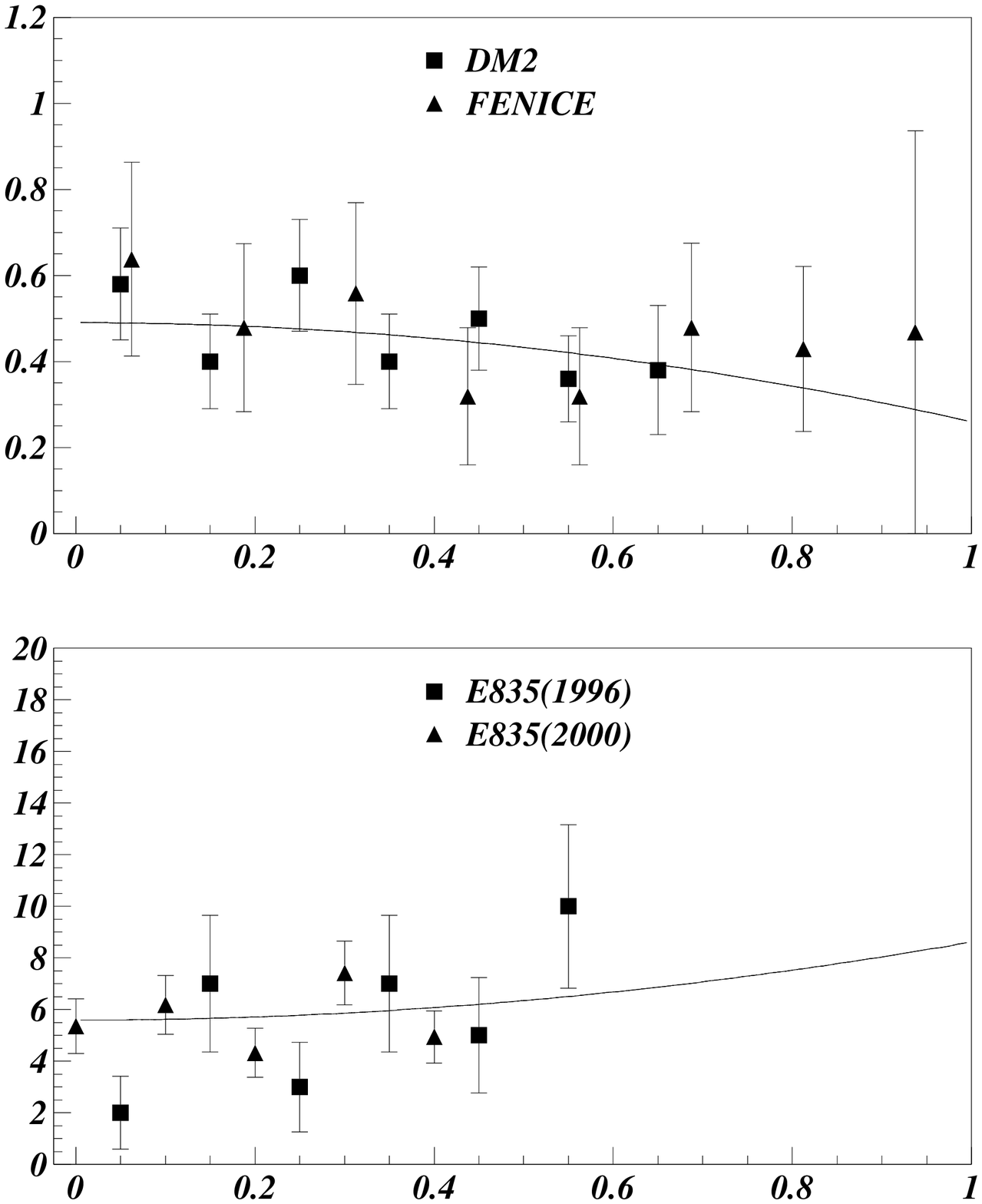,width=100mm}
\put(-247,300){$a$}
\put(-247,144){$b$}
\put(-100,300){$s=4.5\;GeV^2$}
\put(-100,144){$s=12.0\;GeV^2$}
\put(-61,167){$|\cos(\theta)|$}
\put(-61,11){$|\cos(\theta)|$}
\put(-306,255){\rotatebox{90}{\bm{10mm}$$\frac{d\sigma}{d|\cos(\theta)|}(nb)$$\em}}
\put(-306,117){\rotatebox{90}{\bm{10mm}$$\frac{d N}{d|\cos(\theta)|}$$\em}}
\vspace{-5mm}
\caption{a) Differential cross-section as a function of  
    $|\cos(\theta)|$ from FENICE\cite{fenice_1,fenice_2} (triangles)
    and DM2\cite{dm2_1,dm2_2} (full squares) data 
    at $s^{\sf exp}_1=4.5\; GeV^2$, with fit superimposed. 
    b) Angular distributions for E835 1996 data (full squares) and 2000 
    data\cite{andreotti} (triangles) at  $s^{\sf exp}_2=12.0\; GeV^2$, 
    with fit superimposed: the two distributions are 
    normalized to the same number of events per unit of $\cos(\theta)$. 
}
\label{angular}
\end{center}
\end{figure}

The proton ff's in the time-like region ($q^2\equiv s>0$)
can be extracted from the cross-section of the process 
$e^+e^-\rightarrow p\overline{p}$ or 
$p\overline{p}\rightarrow e^+e^-$.
The measurements available from both kinds of experiments are
shown in Fig.\ref{collection} as a function of the center of mass energy 
$s$. 
Most of the data are concentrated in the low $s$ region
close to the proton-antiproton threshold \cite{adone,dm1,dm2_1,dm2_2,apple,fenice_1,fenice_2}, 
but a sizeable amount of data from proton-antiproton annihilation 
experiments is also available for center of mass 
energies $s$ between 8 and 14 $GeV^2$ \cite{e760,e835_1,e835_2}.
The total cross-section at a given $s$ is related to the ff's $|G_E^p(s)|$ 
and $|G_M^p(s)|$ through the relation:
\be
\sigma(s)=\frac{4\pi\alpha^2\rho(s)}{3s}\left(|G_M^p(s)|^2+\frac{2M_p^2}{s}
|G_E^p(s)|^2\right)
\en
where the factor $\rho(s)$ is equal to $C \beta=C v/c$  ($v$ is the proton 
velocity in the center of mass system) in the case of $e^+e^-\rightarrow 
p\overline{p}$ and to $\beta/C$ in the case of 
$p\overline{p}\rightarrow e^+e^-$ and $C$ is a Coulomb correction factor to 
take into account QED bound states, relevant only very near threshold.
In order to extract the ff's from the measured cross-sections, each
experiment has to make an hypothesis on the modulus of the ratio $R(s)$. 
All the results published are obtained in the hypothesis that this ratio,
in modulus, is equal to $\mu_p$ ($|G_E^p(s)|=|G_M^p(s)|$).
\\
The possibility to disentangle between $|G_E^p(s)|$ and
$|G_M^p(s)|$ relies on the measurement of the angular distributions. 
Calling $\theta$ the polar angle of the emerging proton (or antiproton) 
in the center of mass system in $e^+e^-\rightarrow p\overline{p}$ experiments, 
or the polar angle of the electron (or positron) in $p\overline{p}\rightarrow e^+e^-$ 
experiments, the differential cross-section is
\be
\frac{d\sigma}
  {d\Omega}=\frac{\alpha^2\rho(s)}{4s}\left(|G_M^p(s)|^2(1+\cos^2(\theta))+ 
  \frac{1}{\tau}|G_E^p(s)|^2\sin^2(\theta)\right),
\label{rose-tl2} 
\en
where $\tau=s/4M_p^2$.
This formula shows that the two ff's give rise to two terms: one,
related to the magnetic ff, has 
a $[1+\cos^2(\theta)]$ dependence, the other one, related to the electric 
ff has a $\sin^2(\theta)$
dependence. In principle a measurement of the angular distribution is able to
give the relative weights of the two terms and hence the ratio $|R(s)|$.
A fit of the $\cos(\theta)$ distributions has been applied to two sets of data
corresponding to two different values of $s$.
\begin{enumerate}
\item{The data from the FENICE \cite{fenice_2} and DM2 \cite{dm2_2} experiments 
both detecting the process
$e^+e^-\!\!\rightarrow\!\! p\overline{p}$ at an average center of mass energy 
$s=4.51\;GeV^2$ have been simultaneously fitted. The two measured 
differential cross-sections are in good agreement as shown 
in Fig.\ref{angular}a.}
\item{Two sets of data from
the E835 experiment \cite{e835_1,e835_2} detecting the process
$p\overline{p}\rightarrow e^+e^-$ both at $s\sim 12\;GeV^2$ have also been
fitted simultaneously. The angular distributions of the 
two data sets are shown in Fig.\ref{angular}b
normalized to the average number of events per bin.}
\end{enumerate}
\vspace{-0mm}
All the considered experiments are characterized
by limited statistics due to the small value of the cross-section 
(from $\sim1\; nb$ at $s^{\sf exp}_1=4.5 \;GeV^2$ down to 
$\sim 1\; pb$ at $s^{\sf exp}_2=12 \;GeV^2$).
In both cases the parameter  $|R|^{\sf exp}_k$ ($k=1,2$) is 
extracted from a fit to the c.m. angular distribution with:
\be
f_k(\cos\theta)=\tau\mu_p^2A_k^2(1+\cos^2\theta)+B_k^2(1-\cos^2\theta)
\label{fit-tl},
\en
the ratio between the two coefficients $B_k$ and $A_k$ is just 
$|R|^{\sf exp}_k$. In order to evaluate confidence intervals for $|R|^{\sf exp}_k$ 
properly taking into account the unavoidable correlations, 
the same fits have been applied on samples of distributions randomly extracted 
from the experimental ones. This procedure allows to get directly the distributions of 
$|R|^{\sf exp}_k$. These are not gaussian, being in both cases strongly asymmetric. 
A 68\% confidence intervals ($R_k^{inf}<|R|^{\sf exp}_k<R_k^{sup}$) 
are built in such a way that the probability $P(|R|>R_k^{sup})=P(|R|<R_k^{inf})=16\%$.
The resulting intervals are: 
\be
\begin{array}{cc}
 3.46<|R|^{\sf exp}_1 <26.5 & (s^{\sf exp}_1=4.5\; GeV^2)\\
 0.34<|R|^{\sf exp}_2 <3.63 & (s^{\sf exp}_2=12.0\; GeV^2).\\
\end{array}
\label{tl-data}
\en
These numbers could be modified by taking into account the contribution of
two-photon exchange diagrams, that should lead to an asymmetry in the
angular distribution of the proton with respect to the antiproton.
An effect of the order of few percent could be estimated, also taking into
account that the $\gamma \gamma \rightarrow p \bar{p}$ cross section is of
the same order of magnitude as for $ e^+ e^- \rightarrow p \bar{p}$
\cite{gammagamma}.
However, given the experimental uncertainties in the determination of the
two intervals of eq.(\ref{tl-data}), we do not expect that our final result 
will be affected by these higher order corrections.

%%%%%%%%%%%%%%%%%%%%%%%%%%%%%%%%%%%%%%%%%%%%%%%%%%%%%%%%%%%%%%%%%%%%%%%%%%%%%%%%%%%%%%%%%%%%%%%%%%%%%%
\section{The dispersive approach}
\label{sec:approach}

In order to connect the data on $|R(s)|$, in the time-like region to 
those on $R(t)$, in the space-like region ($t=q^2<0$), the 
DR's for the imaginary part is a very powerful mathematical 
tool that we can use. 
The DR for the imaginary part has the form \cite{math}:
\be
G(t)=\frac{1}{\pi}\int_{s_0}^\infty\frac{\im{G(s)}}{s-t}ds.
\label{dr-im}
\en
It establishes an integral relation between the values of the ff 
in the space-like region, where this function is real since 
it describes the scattering process, and its imaginary part in the 
time-like region, where the process described is the annihilation and 
therefore the ff is complex. If we assume that the magnetic proton 
ff $G_M^p(q^2)$ has no zeros, as it is demonstrated in 
ref.\cite{baldini}, it follows that, unless
the asymptotic behaviour, the ratio $R(q^2)$ has the same analytic 
properties of each ff. This means we may apply the
DR of eq.(\ref{dr-im}) directly to $R(q^2)$. 
But, since pQCD\cite{asymp1} constrains the ff's $G_E^p(q^2)$
and $G_M^p(q^2)$ to be asymptotically vanishing with the same 
power law $(1/q^2)^2$, as $q^2$ diverges, the ratio,
unless logarithmic corrections, should have a constant asymptotic 
time-like limit and therefore the integral of eq.(\ref{dr-im}), with
$R(s)$ instead of $G(s)$, could be divergent.
In order to account for this possibility we perform the analytic 
continuation of $R(q^2)$ by means of the DR for the imaginary 
part, subtracted at $t=0$ \cite{math}:
\be
R(t)=R(0)+\frac{t}{\pi}\int_{s_0}^\infty\frac{\im[R(s)]}{s(s-t)}ds,
\label{dr-im-sub}
\en
which relates the space-like value of $R(t)$ to its time-like
imaginary part and:
\be
\re[R(s)]=R(0)+\frac{s}{\pi}\mathcal{P}\!\!\!\int_{s_0}^\infty
\frac{\im[R(s')]}{s'(s'-s)}ds',
\label{re-dr}
\en
 which, instead, connects the real and the imaginary parts of $R(s)$ 
over the cut, i.e., for $s\ge s_0$
($\mathcal{P}$ denotes the principal value).
The price of the subtraction at $t=0$, is the knowledge 
of the value of $R(t)$ at the same point, but, thanks to the
normalization [eq.(\ref{ratio})], in the expressions of the 
eqs.(\ref{dr-im-sub},\ref{re-dr}) we can put $R(0)=1$.
\\
Contrary to the existing models \cite{1/q,logq2,brodsky,iachello,IJL2}, 
which are constructed starting from
the space-like data and only subsequently extended to other energies, 
we start from the imaginary part of the ratio $R(s)$, which is
defined only in the portion of the time-like region over the cut,
and, by means of a rigorous analytic continuation procedure,
we reconstruct the function $R(q^2)$ in the whole $q^2$ complex plane.\\
To parametrize the imaginary part we use a quite general and
model independent form, that is, two series of orthogonal 
Chebyshev polynomials \cite{math}:
\be
\im[R(s)]\equiv I(s)=\left\{
\begin{array}{lll}
\sum_j^M C_j T_j(x)& x=\frac{2s-s_1-s_0}{s_1-s_0} & s_0\le s\le s_1\\
&&\\
\sum_j^N D_j T_j(y)& y=\frac{2s_1}{s}-1            & s>s_1,\\
\end{array}
\right.
\label{im}
\en
with $s\ge s_0$, the two vectors {\boldmath$C$}$=(C_1,C_2,\ldots,C_M)$ and 
{\boldmath$D$}$=(D_1,D_2,\ldots,D_N)$ represent the coefficients and $T_j(x)$ 
is the $j$-th Chebyshev polynomial. The two series in eq.(\ref{im}) cover two 
naturally separated intervals: 
\begin{itemize}
\item the unphysical region $[s_0,s_1]$ with the vector meson resonances,
      where the ff's can not be measured;
\item the experimentally accessible region $(s_1,\infty)$,  
      where the asymptotic regime is attained.
\end{itemize}
%%%%%%%%%%%%%
%%%%%%%%%%%%%
\bfi[h!]\vspace{-5mm}
\bc
\epsfig{file=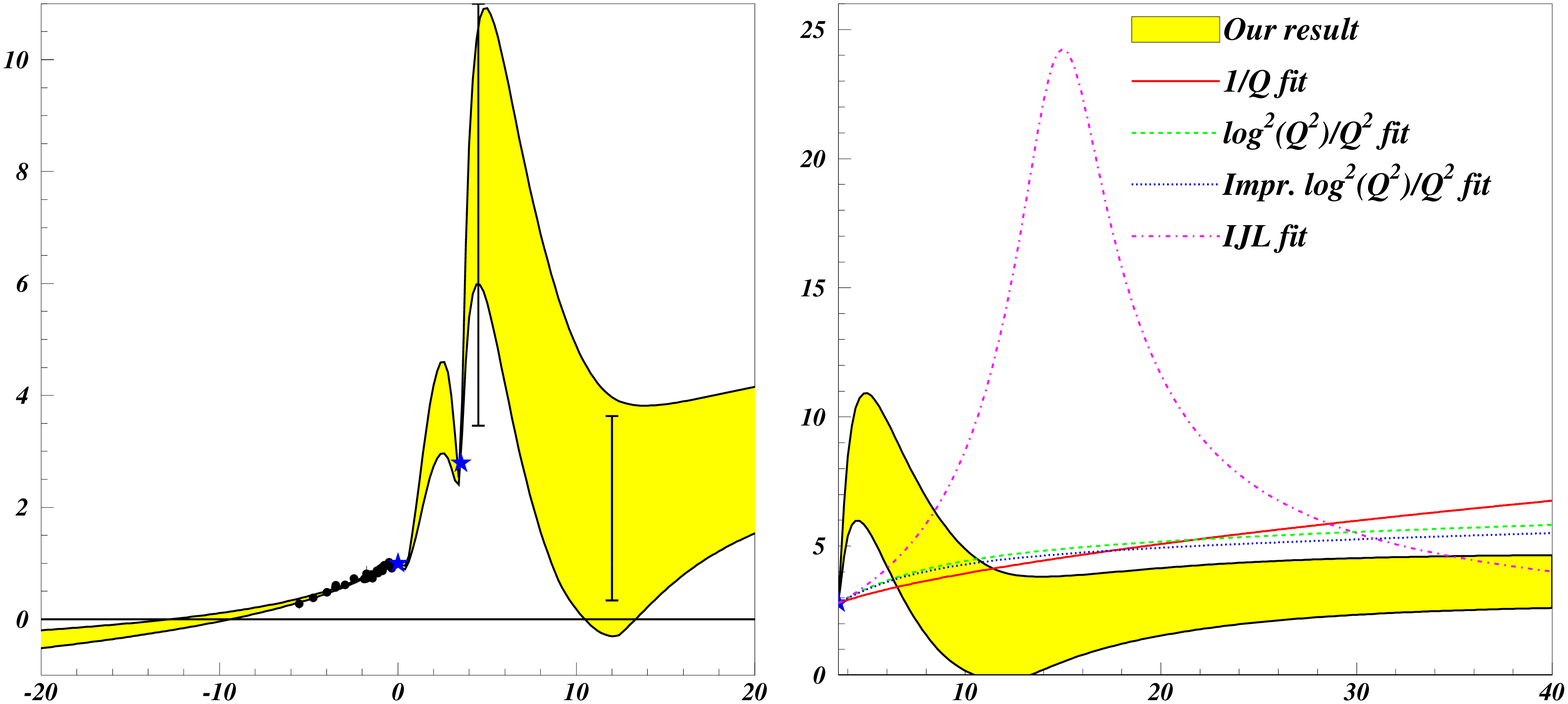,width=120mm}
\put(-193,0){$q^2(GeV^2)$}
\put(-342,125){\rotatebox{90}{$\tilde{R}(q^2)$}}
\put(-156,140){b}
\put(-317,140){a}
\put(-68,135){\scalebox{.6}{\cite{1/q}}}
\put(-47,125){\scalebox{.6}{\cite{logq2}}}
\put(-30,115){\scalebox{.6}{\cite{brodsky}}}
\put(-68,105){\scalebox{.6}{\cite{iachello,IJL2}}}
\caption{a) Result of our dispersive technique for the ratio $R(q^2)$ in the 
full data region. The solid band represents the error. 
b) Ratio $R(q^2)$ in the time-like region with
$q^2\ge s_1$ and comparison among various models ($Q^2\equiv -q^2$). 
As in Fig.\ref{data-fig} the full circles
are the data from Jlab and MIT-Bates, while the two time-like intervals 
are from FENICE, DM2 and E835. The stars represent the theoretical constraints.
The label $\tilde{R}(q^2)$ of the ordinate axis is defined as:
$\tilde{R}(q^2)=R(q^2)$ for $q^2\le s_0$ and $\tilde{R}(q^2)=|R(q^2)|$ for 
$q^2> s_0$.}
\label{r-fig}
\ec
\efi\vspace{-5mm}
%%%%%%%%%%%%%
\bfi[h!]\vspace{-5mm}
\bc
\epsfig{file=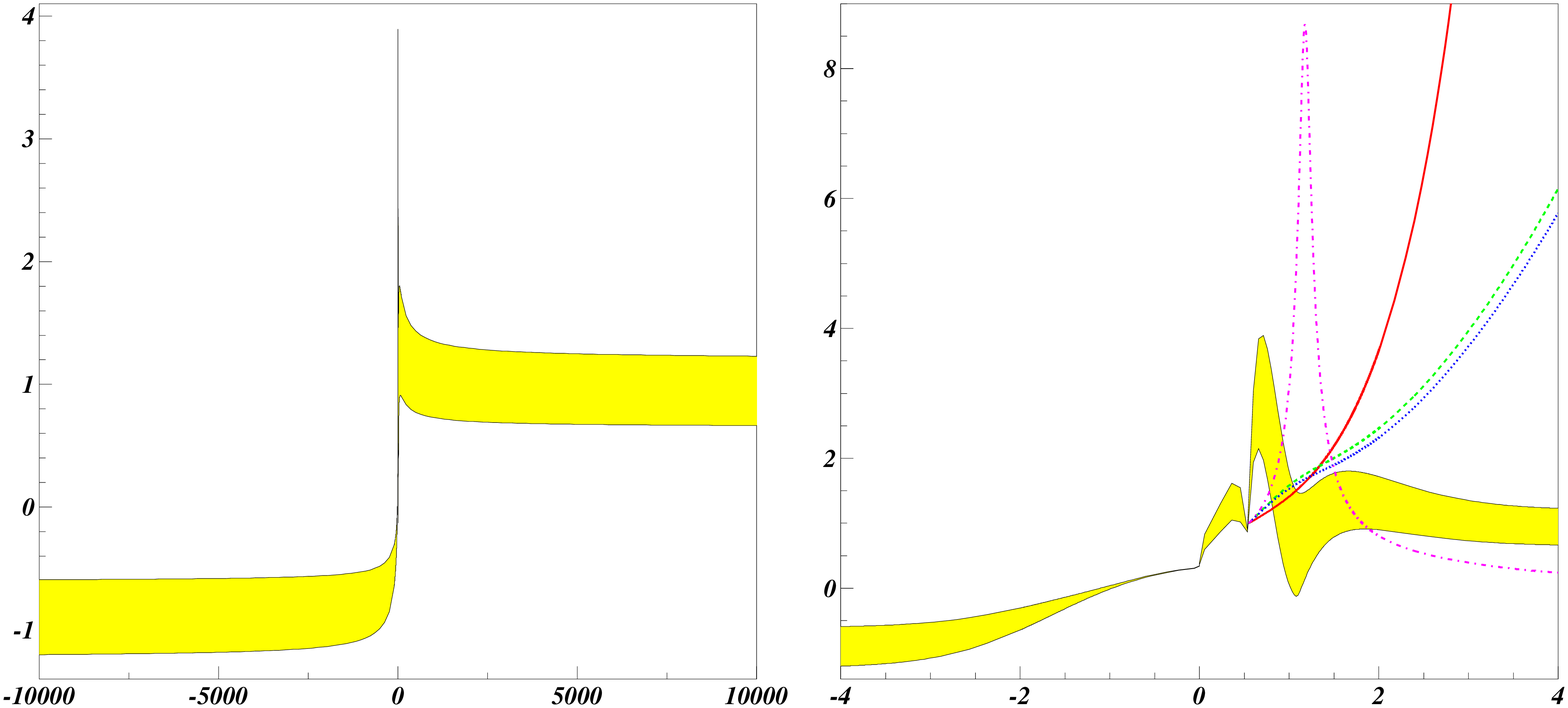,width=120mm}
\put(-77,3){\scalebox{.8}{$\log(q^2/GeV^2)$}}
\put(-156,3){\scalebox{.8}{$-\log(-q^2/GeV^2)$}}
\put(-273,3){\scalebox{.8}{$q^2(GeV^2)$}}
\put(-342,85){\rotatebox{90}{$\tilde{G}^p_E(q^2)/\tilde{G}^p_M(q^2)$}}
\put(-31,140){b}
\put(-192,140){a}
\caption{a) Ratio $G_E^p(q^2)/G_M^p(q^2)$ over a wide range of $q^2$. 
The solid band represents the error. b) Our result compared with the 
other models (same references of Fig.\ref{r-fig}) in logarithmic scale 
for $q^2$. The label $\tilde{G}^p_E(q^2)/\tilde{G}^p_M(q^2)$ of the 
ordinate axis is defined as: 
$\tilde{G}^p_E(q^2)/\tilde{G}^p_M(q^2)=G^p_E(q^2)/G^p_M(q^2)$ 
for $q^2\le s_0$ and 
$\tilde{G}^p_E(q^2)/\tilde{G}^p_M(q^2)=|G^p_E(q^2)/G^p_M(q^2)|$ for 
$q^2> s_0$.}
\label{asy-fig}
\ec
\efi\vspace{-3mm}
%%%%%%%%%%%%%
Once $\im[R(s)]$ is defined, $R(q^2)$ can be evaluated in the full $q^2$ range 
via the eqs.(\ref{dr-im-sub},\ref{re-dr}). In order to find the 
{\boldmath$C$} and {\boldmath$D$} vectors we minimize a $\chi^2$ function 
that includes space-like and time-like data together with the constraints 
discussed above. The resulting function has the correct analytic structure.
More details on the minimization procedure are reported in ref.\cite{simone}.

Concerning the dispersive procedure, it is interesting to mention that,
in addition to the one adopted here, there are other forms of DR's which connect
space-like and time-like data. For instance, the DR for the
logarithm may relate directly data on the modulus of a ff in the time-like 
region to those for its real value in the space-like region \cite{math,baldini,gourdin}.
This kind of logarithmic dispersive approach would not be suitable in this
case, not only because the time-like modulus has not so many constraints as
the imaginary part does, but also because it is not able to forecast,
in a model independent way, the presence of a zero, that, instead, 
should be one of the aims of this analysis.
In the pictures of Fig.\ref{r-fig} is shown our result for the ratio $R(q^2)$.
We found, in a model independent way, a space-like zero at $q^2=(-11\pm 2)\;GeV^2$
(Fig.\ref{r-fig}a). 
The comparison between some existing models\cite{1/q,logq2,brodsky,iachello,IJL2} and
our result, reported in Fig.\ref{r-fig}b, shows that, in spite of the agreement
in the space-like region, where all these models describe the polarization data,
their continuations in the time-like region are far each other. 
Another interesting outcome of our computation is the space-like and time-like 
asymptotic behaviour of $G_E^p(q^2)/G_M^p(q^2)$. According to the
Phragm\`en-Lindel\"off theorem, that we have implemented by imposing
a vanishing asymptotic value for the imaginary part of $R(q^2)$,
we achieved a real time-like limit for the ratio as $q^2\rightarrow+\infty$.
As shown in Fig.\ref{asy-fig} (without and with comparison with the other
models) the space-like and time-like asymptotic limits have, in
modulus the same value, but they have opposite sign. i.e.:
\be
\lim_{q^2\rightarrow \pm \infty}\frac{G_E^p(q^2)}{G_M^p(q^2)}=\pm1.
\label{asy-lim}
\en
The scaling law, providing $G_E^p\simeq G_M^p$, that, in light
of the new polarization data, is no more valid at low $|q^2|$,
has been in some way restored, even if in modulus and in a 
``double'' asymptotic regime. In addition, the 
Phragm\`en-Lindel\"off theorem predicts for the phase
\be
\Phi(\infty)=(Z-P)\pi,
\en
where $Z$ ans $P$ are the number of zeros and poles of $R(q^2)$.
Since we have: 
$R(-\infty) = |R(\infty)|e^{i \Phi(\infty)} = - |R(\infty)|$, and
being $P=0$, we obtain a further confirmation of the presence of a zero of
$R(q^2)$ (or at least an odd number of zeros).
\bfi[h!]\vspace{-3mm}
\bc
\epsfig{file=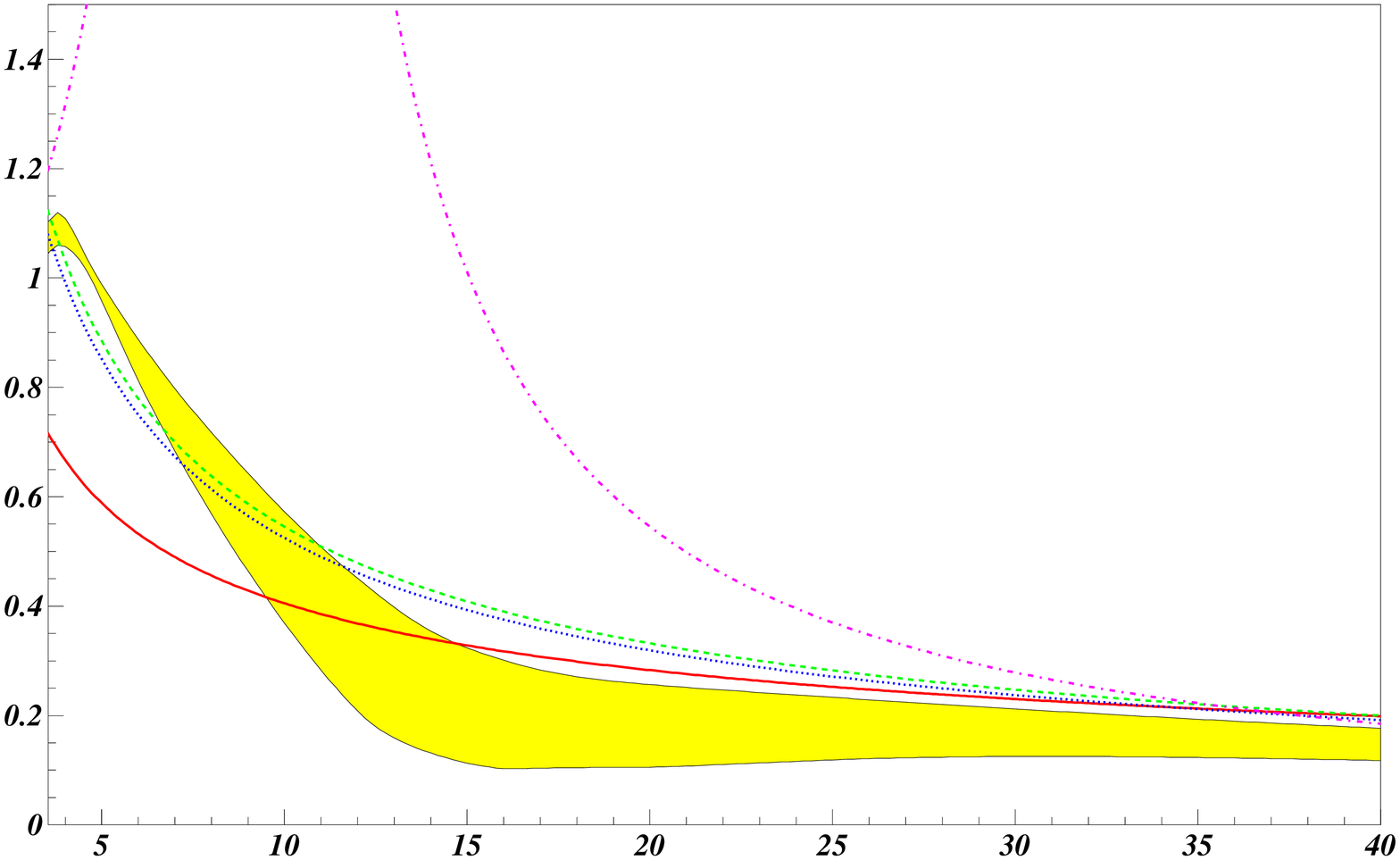,width=100mm}
\put(-50,3){\scalebox{.8}{$q^2(GeV^2)$}}
\put(-287,110){\rotatebox{90}{\scalebox{.9}{$|F^p_2(q^2)/F^p_1(q^2)|$}}}
\vspace{-3mm}
\caption{Ratio $|F_2^p(q^2)/F_1^p(q^2)|$, our prediction (solid band) compared 
with the other models (same references of Fig.\ref{r-fig}).}
\label{asyf2f1-fig}
\ec
\efi\vspace{-3mm}
\\
%%%%%%%%%%%%%%%%%%%%%%%%%%%%%%%%%%%%%%%%%%%%%%%%%%%%%%%%%%%%%%%%%%%%
As a direct consequence of the limit of eq.(\ref{asy-lim}), we
give also a prediction (Fig.\ref{asyf2f1-fig}) for the ratio between 
Pauli and Dirac ff's as:
\be
\lim_{q^2\rightarrow\infty}\tau \frac{F_2^p(q^2)}{F_1^p(q^2)}=-0.2\pm 0.3,
\label{f2f1-asy}
\en
where, as usual, $\tau=\frac{q^2}{4M_p^2}$, therefore 
$F_2^p(q^2)/F_1^p(q^2)$ scales at least like $(1/q^2)$ as $q^2$
diverges.\\
The complete knowledge of the function $R(q^2)$
allows to give predictions concerning also the quantities depending 
explicitly on real and imaginary part (or on modulus and phase) as
polarization observables. In the space-like region, the scattering
of polarized leptons on a proton target gives non trivial polarization
effects even if the target is unpolarized. The polarization of the
outgoing proton, in this case, depends on the product of electric and
magnetic ff's, which in this region are real.\\
In the time-like region, i.e. when we consider the annihilation 
$e^+e^-\rightarrow p\overline{p}$, the complex structure of the 
ff's give rise to special polarization effects: the
outgoing proton may experience a polarization even if 
there are no polarized leptons in the initial state.
This polarization will be determined by the relative phase of $G_E^p(q^2)$ 
and $G_M^p(q^2)$, that, in our case, is just the phase $\Phi(q^2)$
of $R(q^2)$. 
The time-like polarization vector $\vec{\mathcal{P}}$ has the components\cite{dubn}:
\be
\mathcal{P}_y(q^2)\!\!\!&=&\!\!\!-\frac{\sin(2\theta)|R(q^2)|\sin(\Phi(q^2))}
{D\sqrt{\tau}}\nonumber\\
\mathcal{P}_x(q^2)\!\!\!&=&\!\!\!-P_e\frac{ 2\sin(\theta)|R(q^2)|\cos(\Phi(q^2))}
{D\sqrt{\tau}}
\label{pol-sing}
\\
\mathcal{P}_z(q^2)\!\!\!&=&\!\!\!P_e\frac{ 2\cos(\theta)}{D},\nonumber
\en 
with:
\be
D=\frac{1+\cos^2(\theta)+\frac{1}{\tau}|\tau|^2\sin^2(\theta)}{\mu_p},
\en
where $z$ is the direction of the outgoing proton in the center of mass system
and $y$ is orthogonal to the scattering plane, $P_e$ is the longitudinal polarization 
of the initial lepton and $\theta$ is the scattering angle. 
As already said, the $y$-polarization $\mathcal{P}_y$ 
does not depend on $P_e$, while the longitudinal polarization does not
depend on the pahse $\Phi$. In the pictures of Fig.\ref{pxpypz-fig} are
shown our predictions and those of the other considered models\cite{1/q,logq2,brodsky,iachello,IJL2},
for the components of the polarization vector. Once again 
models, which agree in the space-like region, give different 
predictions in the time-like region.
\bfi[h!]\vspace{-5mm}
\bc
\epsfig{file=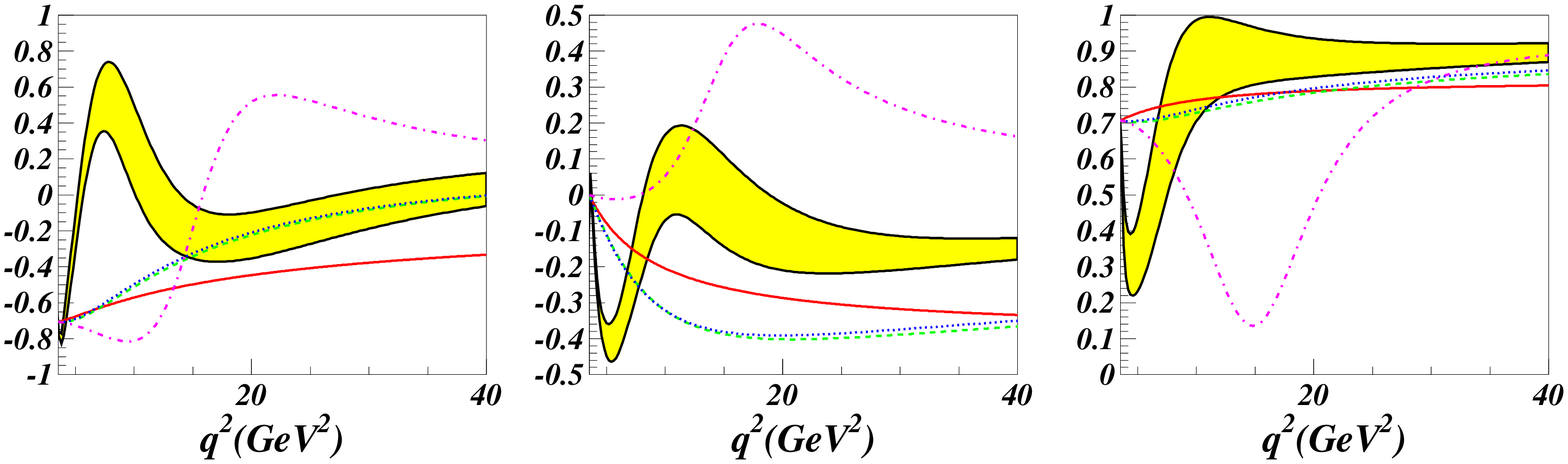,width=160mm}
\put(-378,127){\scalebox{1.}{$\mathcal{P}_x$}}
\put(-236,127){\scalebox{1.}{$\mathcal{P}_y$}}
\put(-94,127){\scalebox{1.}{$\mathcal{P}_z$}}
\vspace{-5mm}
\caption{Our predictions (solid band) for polarizations 
$\mathcal{P}_x$, $\mathcal{P}_y$ and $\mathcal{P}_z$ in the time-like region 
compared with those of the other models (same references of Fig.\ref{r-fig}). 
The plots are for the scattering angle $\theta=45^o$ 
and electron polarization $P_e=1$.}
\label{pxpypz-fig}
\ec
\efi\vspace{-5mm}
%
%
%%%%%%%%%%%%%%%%%%%%%%%%%%%%%%%%%%%%%%%%%%%%%%%%%%%%%%%%%%%%%%%%%%%%%%%%%%%%%%%%%%%%%%%%%%%%%%%%%%%%%%
%
\section{Conclusions and perspectives}
\label{sec:coclu}
We have used a dispersive approach to construct an expression
for the ratio $R(q^2)$, defined in the whole $q^2$ complex plane,
which verifies all the constraints imposed by the theory (analyticity,
asymptotic behaviour, etc.) and by the data available at this moment.
The full knowledge of the function $R(q^2)$ allows to formulate a 
wide range of predictions. In particular, in the time like region,
dominance of the electric form factor near threshold, fading away 
soon, as well as an oscillating pattern is predicted (see Fig.\ref{r-fig}), 
that could be interpreted as a resonances residual, survived to 
the smoothing expected in the ratio. Also the ratio between Dirac and Pauli 
ff's is predicted as well as a definite nucleon polarization. 
The other prediction that should be confirmed or refuted soon, 
is the presence of the space-like zero, that we estimate at 
$q^2=(-11\pm 2)\;GeV^2$, in agreement with ref.\cite{zero-du}. 
New polarization measurements are scheduled at Jlab to push down 
the space-like limit at $q^2\simeq -10\;GeV^2$\cite{new-jlab}.\\
Anyway, the measurement of $|R(q^2)|$ and of the polarizations 
[eq.(\ref{pol-sing})] in the time-like region would have a crucial 
significance not only to disentangle among the models, but also to 
gain a rather complete experimental knowledge of the proton ff's.
%
%
%%%%%%%%%%%%%%%%%%%%%%%%%%%%%%%%%%%%%%%%%%%%%%%%%%%%%%%%%%%%%%%%%%%%%%%%%%%%%%%%%%%%%%%%%%%%%%%%%%%%%%

\section{Acknowledgments}
\label{sec:grazie}

We warmly acknowledge S. Brodsky, S. Dubnicka, F. Iachello, G. Pancheri,
Y. Srivastava for many, important suggestions and discussions. 
We also acknowledge the DM2, E835 and FENICE collaborations for allowing 
partial use of their data. After these paper has been sent for 
pubblication new data have been presented by the BaBar collaboration, 
that corroborate our conclusions. 
%
%
%%%%%%%%%%%%%%%%%%%%%%%%%%%%%%%%%%%%%%%%%%%%%%%%%%%%%%%%%%%%%%%%%%%%%%%%%%%%%%%%%%%%%%%%%%%%%%%%%%%%%%

%
\end{document}